\documentclass[12pt]{iopart}
\usepackage{iopams}  
\usepackage{graphicx}

\newcommand{\mean}[1]{\left\langle {#1} \right\rangle }
\begin{document}

\title{Temperature chaos is a non-local effect}

\author{L.~A.~Fernandez$^{1,2}$, E.~Marinari$^{3,4}$, V.~Martin-Mayor $^{1,2}$, G.~Parisi$^{3,4}$ and D.~Yllanes$^{5,2}$}

\address{$^1$ Depto. de F\'{\i}sica Te\'orica I. Facultad de Ciencias
  F\'{\i}sicas. Universidad Complutense de Madrid. 28040 Madrid. Spain.}
\address{$^2$ Instituto de Biocomputaci\'on y
  F\'{\i}sica de Sistemas Complejos (BIFI), 50018 Zaragoza, Spain.}
\address{$^3$ Dip. di Fisica and INFN--Sezione di
  Roma 1, Universit\`a La Sapienza, P.le A. Moro 2, I-00185 Rome, Italy.}
\address{$^4$ Nanotec-CNR, UOS Roma, Universit\`a La Sapienza, P. le
  A. Moro 2, I-00185, Rome, Italy.}
\address{$^5$ Department of Physics and Soft Matter Program, Syracuse University, Syracuse, NY, 13244, U.S.A.}
\eads{dyllanes@syr.edu}
\date{\today}

\begin{abstract}
  Temperature chaos plays a role in important effects, like for
  example memory and rejuvenation, in spin glasses, colloids,
  polymers. We numerically investigate temperature chaos in spin
  glasses, exploiting its recent characterization as a rare-event
  driven phenomenon. The peculiarities of the transformation from
  periodic to anti-periodic boundary conditions in spin glasses allow
  us to conclude that temperature chaos is non-local: no bounded
  region of the system causes it. We precise the statistical
  relationship between temperature chaos and the free-energy changes
  upon varying boundary conditions.
\end{abstract}

\submitto{Journal of Statistical Mechanics}

\maketitle

\section{Introduction.}
Temperature
chaos~\cite{mckay:82,bray:87b,banavar:87,kondor:89,kondor:93,billoire:00,rizzo:01,mulet:01,billoire:02,krzakala:02,rizzo:03,sasaki:05,katzgraber:07},
is one of the outstanding mysteries posed by spin
glasses~\cite{edwards:75,binder:86,mezard:87,fisher:91,young:98,mezard:09,binder:11b}.
It consists in the complete reorganization of the equilibrium
configurations by the slightest change in temperature.  The topic is
currently under intense theoretical
scrutiny~\cite{parisi:10,fernandez:13,billoire:14,wang:15}, not only
because of its importance to analyze spectacular
experiments~\cite{jonason:98,bellon:00,vincent:00,bouchaud:01b,ozon:03,yardimci:03,mueller:04,guchhait:15b},
but also as a crucial tool to assess the performance of quantum
annealers~\cite{martin-mayor:15,katzgraber:15}.

Here we exploit some of its very peculiar features to show that
temperature chaos is a spatially non-local effect.  For a disordered
system, chaos should be studied on a sample by sample basis. In
particular, for system sizes accessible to equilibrium computer
simulations, chaos is a rare event, present only in a small fraction
of the samples (as the system size increases, so does the fraction of
chaotic samples~\cite{fernandez:13}).  We use this fact by
thermalizing spin glasses down to a very low temperature (well below
the critical temperature $T_\mathrm{c}$).  Then, for each simulated
system, with periodic boundary conditions (PBC), we consider its image
under a transformation where we make the boundary conditions
anti-periodic (APBC) in one direction. As we discuss below, this
transformation amounts to change a tiny fraction of the coupling
constants.  Now, due to the gauge invariance in spin glasses, the
couplings that have been changed by our transformation can be placed
\emph{anywhere} in the lattice.  Interestingly enough, whether or not
the PBC instance is chaotic carries essentially no information on the
behaviour of its APBC transform. It follows that temperature chaos is
not encoded in any localized region of the system.

We remark that our work relates as well to the long-standing
controversy regarding the nature of the spin-glass phase. On the one
hand, the Replica Symmetry Breaking theory (stemming from the
mean-field solution) envisages the spin-glass phase as composed of a
multiplicity of states~\cite{mezard:87,marinari:00}. Thus, from this
point of view, the change of boundary conditions is a strong
perturbation and there are no reasons to expect that temperature chaos
effects will be significantly correlated for the PBC system and its
APBC transform. On the other hand, the droplet
picture~\cite{mcmillan:84,bray:87,fisher:86,fisher:88} expects a
single domain wall difference between the two types of boundary
conditions, so there would be a strong correlation of the temperature
chaos effects for the PBC/APBC systems. In this respect, our data
favour Replica Symmetry Breaking (because little correlation is
observed). However, it has been pointed out many times that resolving
this controversy requires studying much larger systems than it is
accessible to current simulations (or
experiments~\cite{janus:10}). This work is no exception.  Furthermore,
our analysis relies crucially in that the system sizes are modest.
Indeed, we rely in that temperature chaos is a rare-event \emph{on
  small systems} while, for larger systems, one expects that
\emph{typical} samples will display strong chaotic events.

The layout of the remaining part of this paper is as follows. In
Sect.~\ref{sect:model} we recall the model definition and the crucial
quantities we study. Some crucial features of temperature chaos are
presented in Sect.~\ref{sect:T-chaos}. Our main results are given in
Sect.~\ref{sect:results}. We briefly explore the relationships between
the free-energy and temperature chaos in
Sect.~\ref{sect:free-energy}. Our conclusions are given in
Sect.~\ref{sect:conclusions}. Technical details are provided in two
appendices.

\section{The Edwards-Anderson model.}\label{sect:model}
 Our $S_{\boldsymbol u}=\pm 1$ spins
occupy the nodes of a $D=3$ lattice of size $L^3$ endowed with
periodic boundary conditions. The Hamiltonian is
\begin{equation}\label{eq:EA}
\mathcal H = - \sum_{\langle \boldsymbol u ,\boldsymbol v\rangle} S_{\boldsymbol u} J_{\boldsymbol u \boldsymbol v} S_{\boldsymbol v}\, .
\end{equation}
The couplings $J_{\boldsymbol u, \boldsymbol v}$ are $\pm 1$ with
$50\%$ probability and only connect nearest neighbouring sites on the
lattice. A particular realization of these couplings (quenched, i.e.,
fixed once and for all) is called a sample. Thermal averages for fixed
$\{J_{\boldsymbol u,\boldsymbol v}\}$ are denoted by
$\langle\cdots\rangle_J$. This system has a second-order phase
transition at temperature $T_\mathrm{c}=1.102(3)$~\cite{janus:13}.

For any original (periodic, PBC) instance its anti-periodic pair
(APBC) is obtained by reversing the coupling $J_{\boldsymbol
  u,\boldsymbol v}$ that join sites $(x=0,y,z)$ and $(1,y,z)$ for all
values of $y$ and $z$ [only a $1/(3L)$ fraction of the bonds is
  changed~\footnote{In Ref.~\cite{wang:16} \emph{all} the couplings
    undergo a tiny change, which produces a related but different
    bond-chaos effect.}].  The APBC image could be a perfectly
reasonable original instance, and, in fact, it is as probable as its PBC
pair.
\begin{figure}[t]
\hskip25mm\includegraphics[height=0.8\linewidth,angle=270]{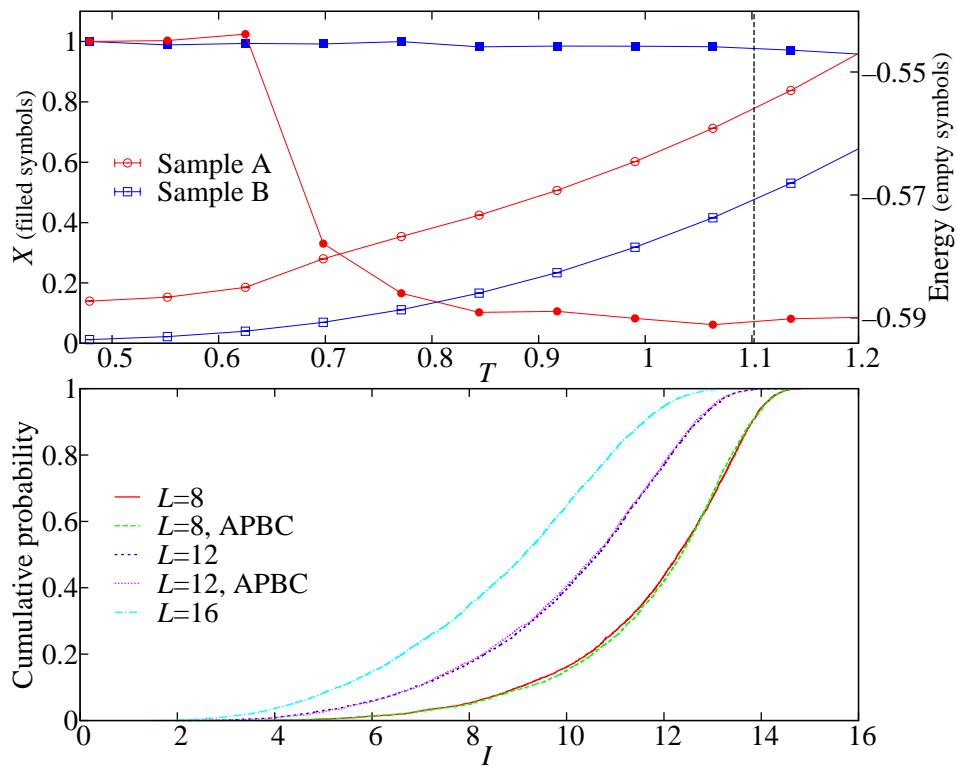}
\caption{\emph{Signatures of temperature chaos.\/} {\bf Top:} for a strongly
  chaotic (A, circles) and a non-chaotic (B, squares) sample, we show the
  chaos parameter $X_{T_\mathrm{min}=0.479,T}$ (solid symbols)
  and the energy per bond $\langle H/(3 L^{3})\rangle_J$ (empty
  symbols) as a function of temperature. The dotted vertical line
  marks the critical temperature. A chaotic event for sample A at
  $T\approx 0.65$ is signalled by the strong drop of $X$. On the
  other hand, the effect in the energy is very subtle.  {\bf Bottom:}
\typeout{XXX is really 0.479 1.6 ?}
  Probability distribution of the chaos integral $I_J$, as computed
  for lattices $L=8,12$ and $16$, with $T_\mathrm{min}=0.479$ and
  $T_\mathrm{max}=1.6$ (for all $L$, the same set of temperatures was
  used in the parallel tempering simulations). The distributions
  obtained with PBC and APBC are, of course, identical. The fraction
  of chaotic samples (i.e., small $I$) increases with system size. $L=16$
  data  from Ref.~\cite{janus:10}.
 \label{fig:chaos}}
\end{figure}

The system described by Eq.~(\ref{eq:EA}) has a gauge
invariance~\cite{toulouse:77}. The energy remains unchanged under the
following transformation:
\begin{equation}
  J_{\boldsymbol u,\boldsymbol v} \longrightarrow \epsilon_{\boldsymbol u}
  \epsilon_{\boldsymbol v} J_{\boldsymbol u,\boldsymbol v}\,, \quad
  S_{\boldsymbol u}\longrightarrow \epsilon_{\boldsymbol u}
  S_{\boldsymbol u}\;, \label{eq:gauge}
\end{equation} 
where $\epsilon_{\boldsymbol u}=\pm1$ can be chosen arbitrarily for
each site $\boldsymbol u$.  Now consider the transformation where
$\epsilon_{(1,y,z)}=-1$ and all other $\epsilon_{\boldsymbol u}=1$.
This changes only the $J_{\boldsymbol u,\boldsymbol v}$ that were
reversed by the APBC transformation and those joining planes $x=1$ and
$x=2$, moving in this way  the transformed-couplings plane from $x=0$ to
$x=1$. Using the same idea, we can place the transformed plane at any
$x$. Furthermore, one can deform the plane of inverted couplings locally in an
essentially arbitrary way by considering a more complicated gauge
transformation. In short, the PBC $\leftrightarrow$ APBC
transformation is non-local.

Another consequence of this gauge symmetry is the need to use real replicas
of the system (i.e., copies that evolve independently but share the same
couplings) in order to form gauge-invariant observables 
(see, e.g.,~\cite{janus:10}).
\begin{figure}[t]
\begin{center}
\includegraphics[height=0.8\linewidth,angle=270]{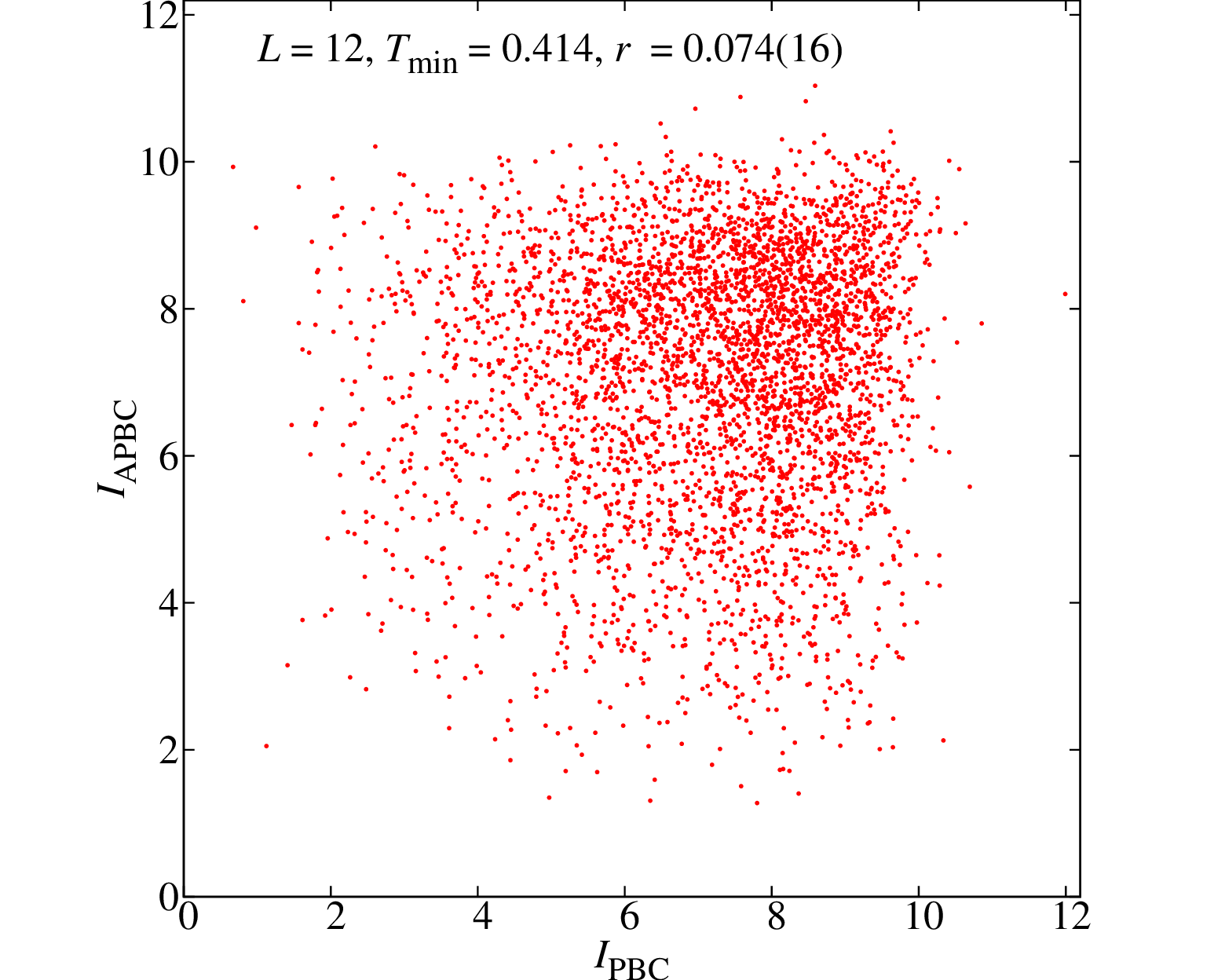}
\end{center}
\caption{\emph{The non-local nature of chaos.\/} Scatter plot of the
  chaos integral $(I_\mathrm{PBC},I_\mathrm{APBC})$ as computed for
  each of our 4000 pairs of samples with $L=12$ and
  $T_\mathrm{min}=0.414$. Hereafter we shall only show data from this
  simulation (the qualitatively identical results for other values of
  $L$ and $T_\mathrm{min}$ can be found in~\ref{ap:additional}).
\label{fig:scatter-I}}
\end{figure}

We have simulated system sizes $L=8,12$ with parallel
tempering~\cite{hukushima:96,marinari:98b}, carrying out several sets
of runs for varying minimum temperature:
$T_\mathrm{min}=0.15,0.414,0.479$ for $L=8$ and $T_\mathrm{min}=
0.414,0.479$ for $L=12$. We have studied the same $4000$ samples and
their $4000$ APBC counterparts for all $T_\mathrm{min}$. Since we want
to study single-sample quantities and chaos, it has been very
important to assess thermalization sample by sample by studying the
temperature-mixing auto-correlation time of the parallel
tempering~\cite{fernandez:09b}. In particular, we use the
thermalization criteria of~\cite{janus:10} (see also~\ref{ap:parameters}).
\section{Some crucial facts about temperature chaos.}\label{sect:T-chaos}
Recently there has been much progress in the numerical
characterization of temperature
chaos~\cite{fernandez:13,billoire:14,martin-mayor:15}.  A
distinguishing feature of a chaotic sample is a very long
auto-correlation time $\tau$ for temperature-mixing along a parallel
tempering simulation. Unfortunately, $\tau$ is very difficult to
measure with any precision, even for well equilibrated
systems~\cite{janus:10}.  As shown in~\cite{fernandez:13}, see
also~\ref{ap:I-or-tau}, this difficulty can be skirted by choosing a
different quantity, easier to measure but strongly correlated with
$\tau$.

In particular, we study the overlap between the spin configurations at
temperatures $T_1$ and $T_2$,
\begin{equation}\label{eq:overlap}
q^{}_{T_1,T_2}=\frac{1}{L^3}\sum_{\boldsymbol{u}}   q_{\boldsymbol{u}}^{T_1,T_2}\,,\quad \mathrm{with}\quad  q_{\boldsymbol{u}}^{T_1,T_2}= S_{\boldsymbol{u}}^{T_1} S_{\boldsymbol{u}}^{T_2}\,,
\end{equation}
and use it to define a {\em chaos parameter}~\cite{ney-nifle:97}:
\begin{eqnarray} 
X_{T_1,T_2}&=\mean{q^2_{T_1,T_2}}_J\big/\big(\mean{q^2_{T_1,T_1}}_J\mean{q^2_{T_2,T_2}}_J\big)^{1/2}\,, \label{eq:X} \\
I &= \sum_{T_2} X_{T_\mathrm{min}, T_2}\, . \label{eq:I}
\end{eqnarray}
In these equations, $\{S^{T_1}\}$ and $\{S^{T_2}\}$ are extracted from
different real replicas. $X_{T_1,T_2}$ is small when the equilibrium
configurations at $T_1$ and $T_2$ differ
significantly. Instead, $X_{T_1,T_2}\approx 1$ in the absence of
temperature chaos.
\begin{figure}[t]
\begin{center}
\includegraphics[height=0.7\linewidth,angle=270]{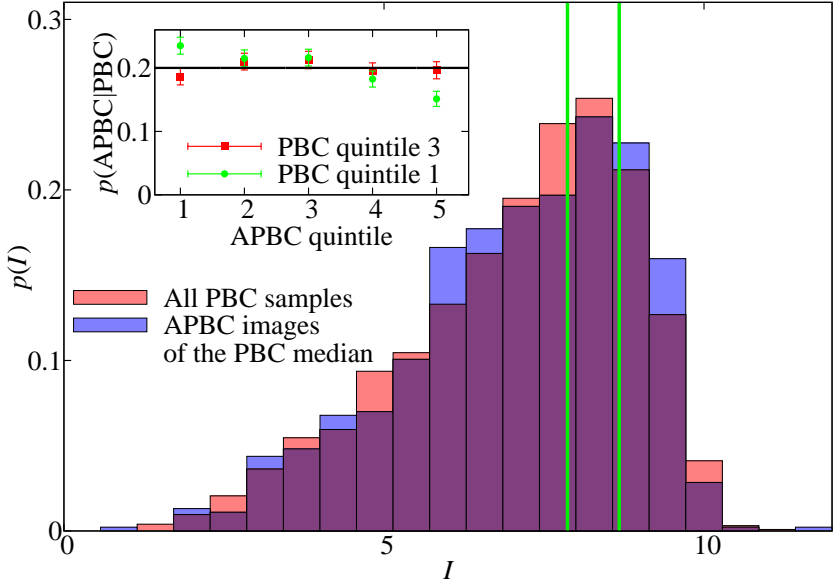}
\end{center}
\caption{\emph{Absence of correlations.\/} The red histogram is the
  probability density function of the chaos integral $p(I)$ for our
  PBC samples ($L=12$, $T_\mathrm{min}=0.414$). Now, let us consider
  the typical samples, namely those spanning the $20\%$ probability
  range around the median (the third quintile, if we divide the population
  in fifths). These third quintile samples 
  belong to the narrow $I$ interval bounded by the two vertical
  green lines. Since temperature chaos is still rare for this system
  size, the third quintile samples are not chaotic.
  However, their APBC transforms (blue histogram) span the
  whole $I$ interval and reproduce the $p(I)$ for PBC. Indeed, an
  Anderson-Darling non-parametric test~\cite{scholz:87} yields a
  $p$-value of $53\%$ for the equal-distribution hypothesis. {\bf Inset:}
  had we selected the $20\%$ most chaotic PBC instances (first
  quintile) the $p(I)$ for their APBC images would be slightly but
  definitively biased towards small $I_J$ (Anderson-Darling
  $p$-value of $0.08\%$). Indeed, the inset shows the conditional
  probability of having $I_{APBC}$ in the $k$-th quintile, given a
  fixed PBC quintile. For the central PBC quintile, the APBC
  conditional probability is uniform. For the first PBC quintile the
  APBC conditional probability is very slightly weighted to small
  $I$.
 \label{fig:quintiles}}
\end{figure}

\begin{figure}[t]
\begin{center}
\includegraphics[height=0.65\linewidth,angle=270]{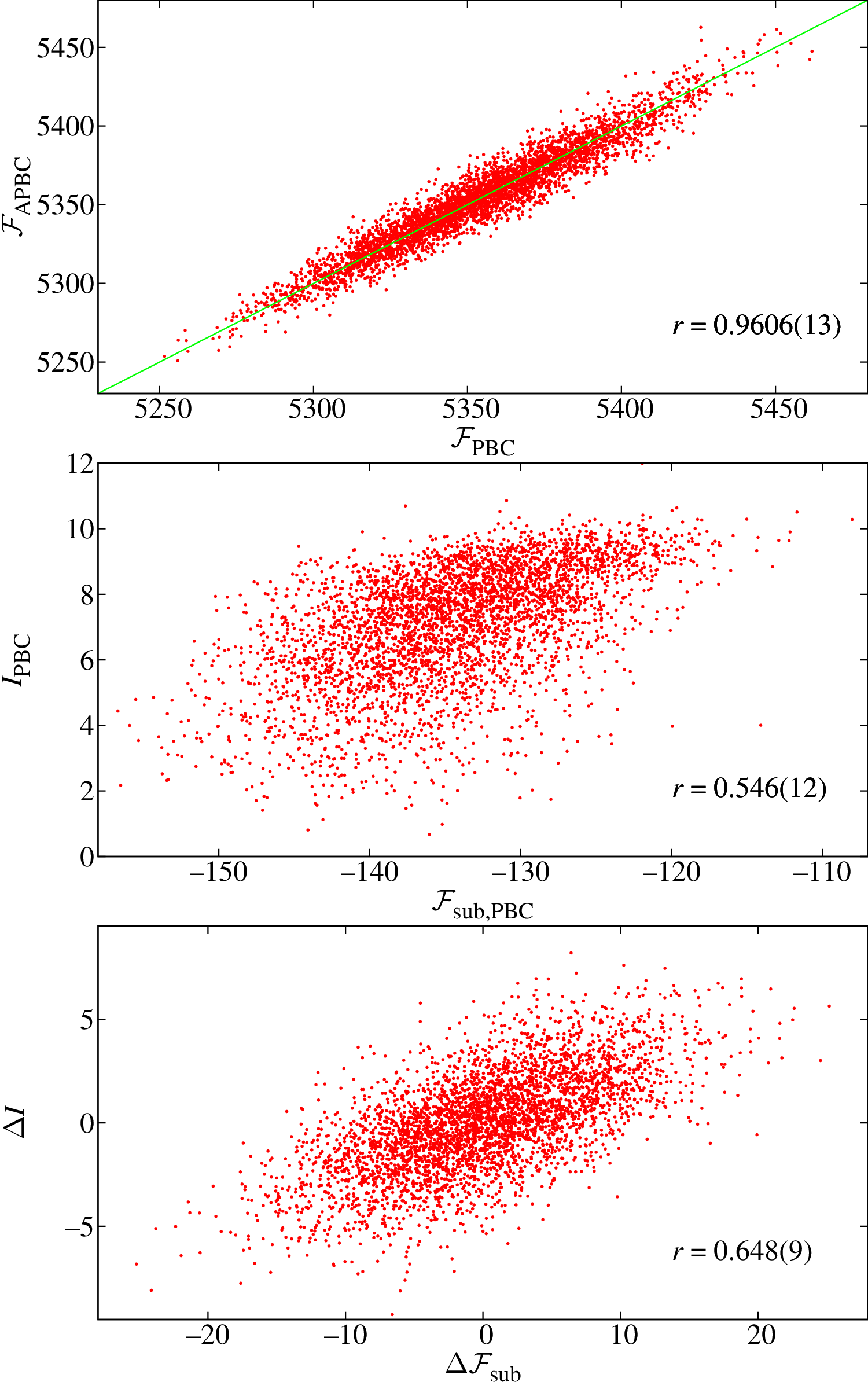}
\end{center}
\caption{\emph{Temperature chaos and the free energy.\/} We show
several scatter plots for our $L=12$ and $T_{\mathrm{min}}=0.414$ data.
{\bf Top:} For the free-energy~(\ref{eq:F}) we plot $({\cal F}_\mathrm{PBC},{\cal F}_\mathrm{APBC})$ for each of our
4000 sample pairs. The green straight line is $y=x$. {\bf Center:} For the subtracted free-energy~(\ref{eq:F-sub}) we plot $({\cal F}_\mathrm{sub,PBC},I_\mathrm{PBC})$. {\bf Bottom:}  we plot
$({\cal F}_\mathrm{sub,PBC}-{\cal F}_\mathrm{sub,APBC},I_\mathrm{PBC}-I_\mathrm{APBC})$.
 \label{fig:F-I}}
\end{figure}

We illustrate the ideas behind these parameters in
Figure~\ref{fig:chaos}---top, where we represent
$X_{T_\mathrm{min},T_2}$ for two samples $A$ and $B$. The ratio of
their respective temperature-mixing times is
$\tau^\mathrm{A}/\tau^\mathrm{B}\approx 3000$. Consequently, for sample A
we can appreciate a very sudden drop in $X_{T_\mathrm{min},T_2}$ (which
we name a chaotic event) for a low value of $T_2$, while sample B has
a smooth $X$. This behaviour can be summarized by saying that chaotic
samples (such as A) have a low value of $I$ (essentially the
integral of $X$), while non-chaotic samples have a high $I$.
Figure~\ref{fig:chaos}---bottom shows that the probability of finding a
chaotic event in a prefixed temperature interval increases for larger
system size~\footnote{For fixed $\epsilon>0$, $T_1$ and $T_2$ the probability of
  having $X_{T_1,T_2}>\epsilon$ drops exponentially in
  $L^3$~\cite{parisi:10,fernandez:13}.}. 

We note also from Figure~\ref{fig:chaos}---top that the energy relates
only in a very subtle way with temperature chaos. We shall further
explore this relation below, since this quantity has been much emphasized 
in the literature~\cite{fisher:88b,sasaki:05,katzgraber:07,wang:15}.

Finally, let us mention that we will base our analysis on
$I$. However, essentially identical results are obtained from the
temperature-mixing time $\tau$, as shown in~\ref{ap:I-or-tau}.

\section{Results.}\label{sect:results}
As Fig.~\ref{fig:scatter-I} shows, the chaos integral $I$ for a
sample has very low correlation to the $I$ value for its APBC
transform.  Yet, the $I$'s are not normally distributed (because the
shape of the scatter plot is not elliptical). As a consequence, the
characterization of correlations through the \emph{very} small
correlation parameter $r=0.074(16)$ is not complete. However, we can
confirm the virtual absence of correlations is confirmed by a more refined
analysis.

We start by considering the full probability distribution
of $I$ for our PBC samples, which spans a wide range 
of values and of course coincides with the $p(I)$
for all the APBC samples. We then take the most typical samples,
those contained in an interval of 20\% probability
around the median (the third quintile). All these samples,
which span a very narrow $I$ range, are non-chaotic. If we
consider the APBC transforms of this median samples
at a first glance one can observe that they span the 
whole $I$ and therefore contain also chaotic instances.
More precisely, one can construct the histogram
of $I$ values for the APBC images of the PBC median,
which turns out to reproduce exactly the full probability
distribution of $I$ for this system.  This is graphically 
shown in Fig.~\ref{fig:quintiles} but it can be proven 
using statistical methods. In particular, an Anderson-Darling
non-parametric test~\cite{scholz:87} finds no 
difference between the full probability distribution of $I$
for the $L=12, T_\mathrm{min}=0.414$ system and the probability
distribution of the images of the (non-chaotic) PBC median.

In short, the $I$ value of the APBC image of a median sample
is completely uncorrelated with its $I_\mathrm{PBC}$. If we repeat
the same analysis, using not the median PBC samples but 
the 20\% most chaotic ones (the first quintile in $I$) we would
find that again the APBC images span the whole $I$ range
but now with a small bias toward low $I$ (see inset to Fig.~\ref{fig:quintiles}).
This is the reason for the non-Gaussian behaviour observed in Fig.~\ref{fig:scatter-I}.

Note that Figs.~\ref{fig:scatter-I} and~\ref{fig:quintiles} are
obtained from only one of our sets of simulations, but our results are
essentially $L$- and $T_\mathrm{min}$-independent (see~\ref{ap:additional}).

\section{Free energy and temperature chaos.}\label{sect:free-energy}
The free-energy change upon varying boundary conditions, $\Delta F =
F_\mathrm{APBC} -F_\mathrm{PBC}$, has received much
attention~\cite{fisher:88b,sasaki:05,katzgraber:07,wang:15}. However,
to the best of our knowledge, the relation between $\Delta F$ and the
spin correlations [e.g., the chaos parameter $X$~(\ref{eq:X})], is
yet to be researched. We can investigate $\Delta F$ from our parallel
tempering simulations by means of thermodynamic integration:
\begin{equation}\label{eq:F}
{\cal F}= \frac{F(T_\mathrm{min})}{T_\mathrm{min}} - \frac{F(T_\mathrm{max})}{T_\mathrm{max}}=\int_{T_\mathrm{min}}^{T_\mathrm{max}}\mathrm{d}\,T\,\frac{\langle H\rangle_{J,T}}{T^2}\,.
\end{equation}
Here, $T_\mathrm{max}=1.6$ is the maximum temperature in our parallel
tempering simulation.  Note that, for large enough $T_\mathrm{max}$,
$\Delta{\cal F}=\Delta F/T_\mathrm{min}$. Indeed, for a
temperature $T$ such that the high-temperature expansion
converges~\cite{parisi:88}, $\Delta F(T)$ goes to zero exponentially
in $L$.

Figure~\ref{fig:chaos}---top shows that chaotic events have an
impact, albeit subtle, on the temperature evolution of $\langle
H\rangle_{J,T}$. Hence, we expect some correlation between $X$ and the
free energy.  The question we address here is: how can we
extract these correlations?

First, we note that $\mathcal{F}$ is not a good chaos indicator
by itself. This is clear already from Fig.~\ref{fig:chaos},
but can be be seen more explicitly in Figure~\ref{fig:F-I}---top, where we
show that $\mathcal{F}_\mathrm{PBC}$ and $\mathcal{F}_\mathrm{APBC}$ 
are almost equal (their correlation parameter is about $r\approx0.95$).
However, on a closer inspection one realizes that chaotic events,
even close to $T=0$, result in minimal  energy changes~\cite{boettcher:04b}.
In other words, even in the most favourable case where only one member of the 
(PBC, APBC) pair has a chaotic event, the energy difference between 
the two samples is very small.  Therefore, some sort of background 
subtraction is needed to enhance the chaotic signal:
\begin{equation}\label{eq:F-sub}
{\cal F}_{\mathrm{sub}}= \int_{T_\mathrm{max}}^{T_\mathrm{min}}\mathrm{d}\,T\,\frac{\langle H\rangle_{J,T} - \langle H\rangle_{J,T_\mathrm{min}}}{T^2}
\end{equation}
Notice that, at very low temperature, (\ref{eq:F-sub}) highlights 
the entropic contribution. 

As we in see Figure~\ref{fig:F-I}---centre,
$\mathcal{F}_\mathrm{sub}$ is correlated with the chaos integral $I$. However, 
this correlation is only of $r=0.546(12)$ and the scatter plot has a 
non-trivial structure, seemingly composed of two different
populations. Therefore, $\mathcal F_\mathrm{sub}$, by itself, still does not seem
a very good indicator of temperature chaos.

It is important to notice that the very strong correlation
of $(\mathcal F_\mathrm{PBC},\mathcal F_\mathrm{APBC})$,
together with the (weaker) correlation of $(\mathcal F_\mathrm{PBC}, I_\mathrm{PBC})$
is in no contradiction with our previous assertion that
$(I_\mathrm{PBC},I_\mathrm{APBC})$ are uncorrelated.

In order to see why, let us consider two stochastic variables, $A$ and $B$,
that have the same variance $V$ and covariance $C$. Their covariance matrix has
eigenvalues $\lambda_\pm=V\pm C$, and the corresponding normal coordinates are
$N_\pm=(A\pm B)/\sqrt{2}$. It follows that the correlation coefficient is
\begin{equation}\label{eq:r}
r = \frac{C}{V} =\frac{1-\frac{\lambda_-}{\lambda_+}}{1+\frac{\lambda_-}{\lambda_+}}\,.
\end{equation}
Clearly, $F_\mathrm{APBC}$ and $F_\mathrm{PBC}$, play the role of the
stochastic variables $A$ and $B$ in the above reasoning. Now, as we shall
explain next, there is a physical reason implying that $\lambda_+$ is orders of
magnitude larger than $\lambda_-$. It follows that the correlation coefficient
is $r\approx 1$ (as we find indeed). In other words, the only information
in Fig. 4-top is $\lambda_+\gg\lambda_-$. 

The physical reason underlying $\lambda_+\gg\lambda_-$ is quite
simple. On the one hand the sample to sample 
fluctuations of the free-energy are of order $L^{D/2}$.
On the other hand, $\lambda_-$, which is the
variance of $(\mathcal{F}_\mathrm{APBC} -\mathcal{F}_\mathrm{PBC})/\sqrt{2}$ scales with
the so called stiffness exponent $\lambda_- \propto L^{2y}$
with $y\approx0.24$ in $D=3$~\cite{boettcher:05}.
An elementary computation
tells us that $\lambda_-/\lambda_+\propto L^{-x}$ with $x=D-2y\approx 2$.

A consequence of this analysis is that the free-energy difference $\Delta \mathcal
F_\mathrm{sub}=(\mathcal F_\mathrm{sub, APBC} - \mathcal F_\mathrm{sub,
PBC})$ is probably a much better chaos indicator than $\mathcal
F_\mathrm{sub}$ by itself.  This is confirmed by Figure~\ref{fig:F-I}---bottom,
which shows an enhanced correlation between $\Delta \mathcal F_\mathrm{sub}$ 
and $\Delta I$, with a more Gaussian behaviour.

In conclusion, the free energy is related to temperature chaos
as studied from the spatial correlation functions, but its
sample-to-sample fluctuations are affected by several
factors not related to chaos (see~\ref{ap:ineq}). Therefore, a refined analysis
is needed to extract information about the chaos integrals from $F$.

\section{Conclusions.}\label{sect:conclusions}
We have shown that temperature chaos, one of the most complex effects
in glass physics, is a non-local phenomenon.  Our approach has two
fundamental ingredients: the recent rare-event characterization of
chaos and the very special nature of periodic boundary conditions
transformations in disordered systems. In fact, anti-periodic boundary
conditions cannot be precisely located in a finite region of the
system. So, changing a tiny [$\mathcal O(1/L)$] fraction of the
coupling constants produces a dramatic effect in the physics of the
considered sample (and the spatial location of the changed couplings
has little importance).

\section*{Acknowledgments}
We thank W. Kob for calling our attention to this problem.  This work
was partially supported by MINECO (Spain) through Grant Nos. FIS2012-35719-C02, FIS2015-65078-C2-1-P. DY
acknowledges support by NSF-DMR-305184 and by the Soft Matter Program
at Syracuse University. Our simulations were carried out on the
Memento supercomputer. We thankfully acknowledge the resources,
technical expertise and assistance provided by BIFI-ZCAM (Universidad
de Zaragoza).

\appendix
\section{Simulation parameters}\label{ap:parameters}
\begin{table}
\centering
\caption{Parameters of our parallel-tempering simulations. In all
  cases we have simulated four independent real replicas for
each of our $4000$  samples. The $N_T$ temperatures are uniformly distributed between
  $T_\mathrm{min}$ and $T_\mathrm{max}$. In this table
  $N_\mathrm{mes}$ is the number of heat-bath sweeps between
  measurements (we perform one parallel-tempering update every 10
  heat-bath sweeps). The simulation length was adapted to the
  thermalization time of each sample (see~\cite{janus:10}). The table
  shows the minimum, maximum and medium simulation times
  ($N_\mathrm{HB}$) for each lattice, in heat-bath steps.}
\label{tab:parameters}
\begin{tabular*}{\columnwidth}{@{\extracolsep{\fill}}ccccrrrr}
\br
 $L$ &  $T_{\mathrm{min}}$ &  $T_{\mathrm{max}}$ &
  $N_T$ &  $N_{\mathrm{mes}}$ &  $N_\mathrm{HB}^{\mathrm{min}}$ &
\multicolumn{1}{c}{ $N_\mathrm{HB}^\mathrm{max}$}& \multicolumn{1}{c}{$N_\mathrm{HB}^\mathrm{med}$}\\
\mr
8  & 0.150  & 1.575 & 10 & $10^3$         & $5\times 10^6$ & $8.30\times 10^8$   & $7.82\times 10^6$\\
8  & 0.414 & 1.554 & 10 & $10^3$         & $10^7$         & $2.00\times 10^7$   & $1.00\times 10^7$\\
8  & 0.479 & 1.619 &  7 & $10^3$         & $10^7$         & $10^7$              & $10^7$           \\
8  & 0.479 & 1.575 & 16 & $10^3$         & $10^7$         & $10^7$              & $10^7$           \\
12 & 0.414 & 1.575 & 12 & $5\times 10^3$ & $10^7$         & $1.53\times 10^{10}$ & $4.60\times 10^7$\\
12 & 0.479 & 1.640 & 12 & $2\times 10^3$ & $8\times 10^6$ & $7.49\times 10^8$    & $1.08\times 10^7$\\
12 & 0.479 & 1.575 & 16 & $2\times 10^3$ & $8\times 10^6$ & $2.56\times 10^9$    & $1.94\times 10^7$\\
\br
\end{tabular*}
\end{table}

Our parallel tempering simulations closely follow
Ref.~\cite{janus:10}.  Some details are provided in
Table~\ref{tab:parameters} for the sake of completeness. There are two
simulation phases. In the first phase, all the PBC instances (and
their APBC images) are simulated for the same amount of time (which is
referred to in Table~\ref{tab:parameters} as the minimum simulation
time $N_\mathrm{HB}^\mathrm{min}$). At that point, we attempt a first
estimate of the temperature-mixing time $\tau$ for each instance and
check that the thermalization criteria were met~\cite{janus:10}. We
chose $N_\mathrm{HB}^\mathrm{min}$ in such a way that most instances
(at least a $2/3$ fraction) are well thermalized. For the remaining
instances, the simulation length is increased and $\tau$
recomputed. The procedure follows until safe thermalization is
achieved.

Some of the simulations for our PBC-samples were actually taken from
Ref.~\cite{janus:10}, specifically the $(L=8,T_\mathrm{min}=0.15)$ and
$(L=12,T_\mathrm{min}=0.414)$ simulations. We did perform totally new
simulations for the APBC image of this system. Additional simulations
were performed in order to show the size-dependence in
Fig.~\ref{fig:chaos}---bottom (the comparison of the chaos integral is
easiest if we employ the same temperature grid in the Parallel
Tempering for all system sizes).

\section{Mixing time or chaos integral?}\label{ap:I-or-tau}

\begin{figure}[h]
\begin{minipage}[t]{\linewidth}
\includegraphics[height=0.525\linewidth,angle=270]{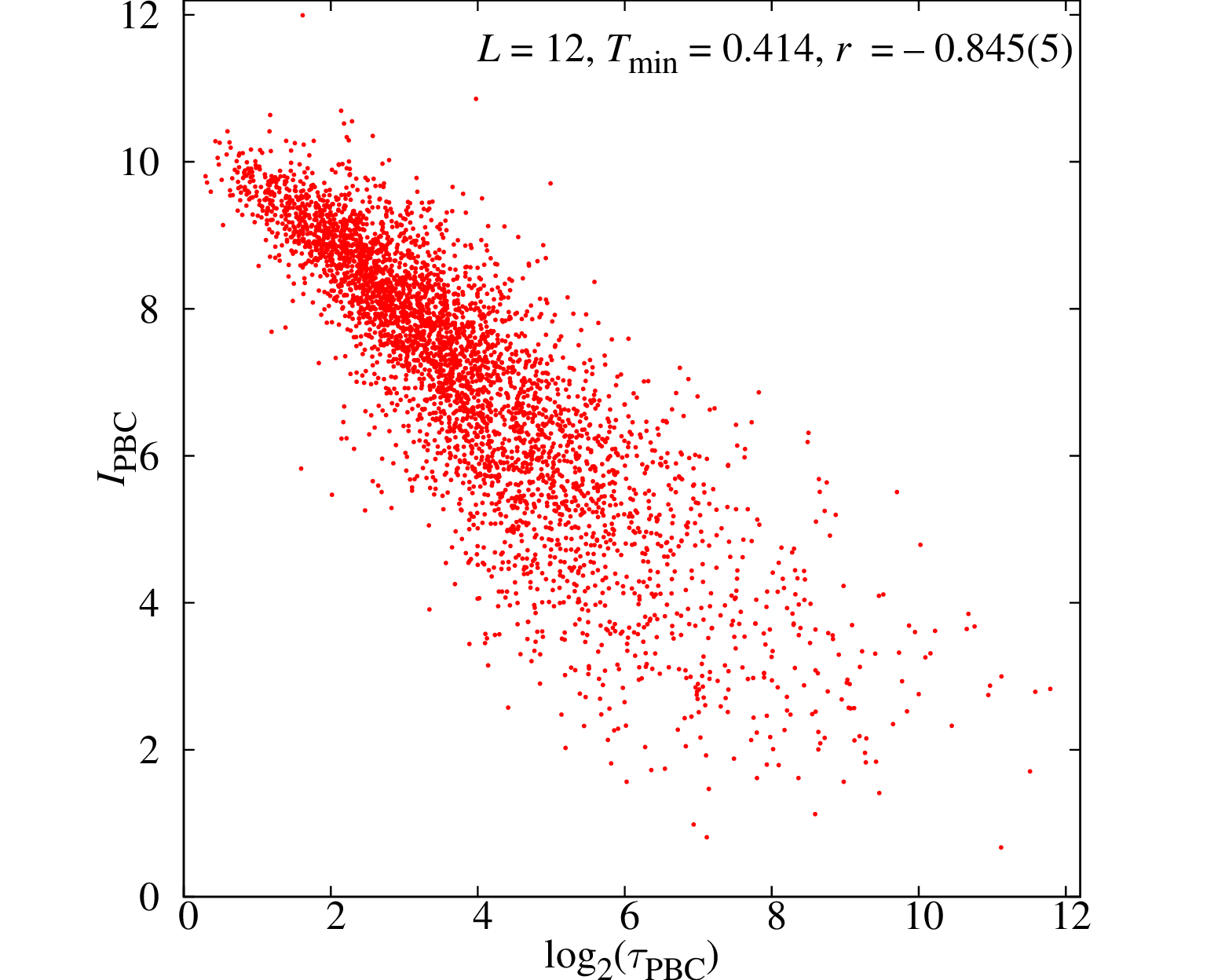}
\includegraphics[height=0.525\linewidth,angle=270]{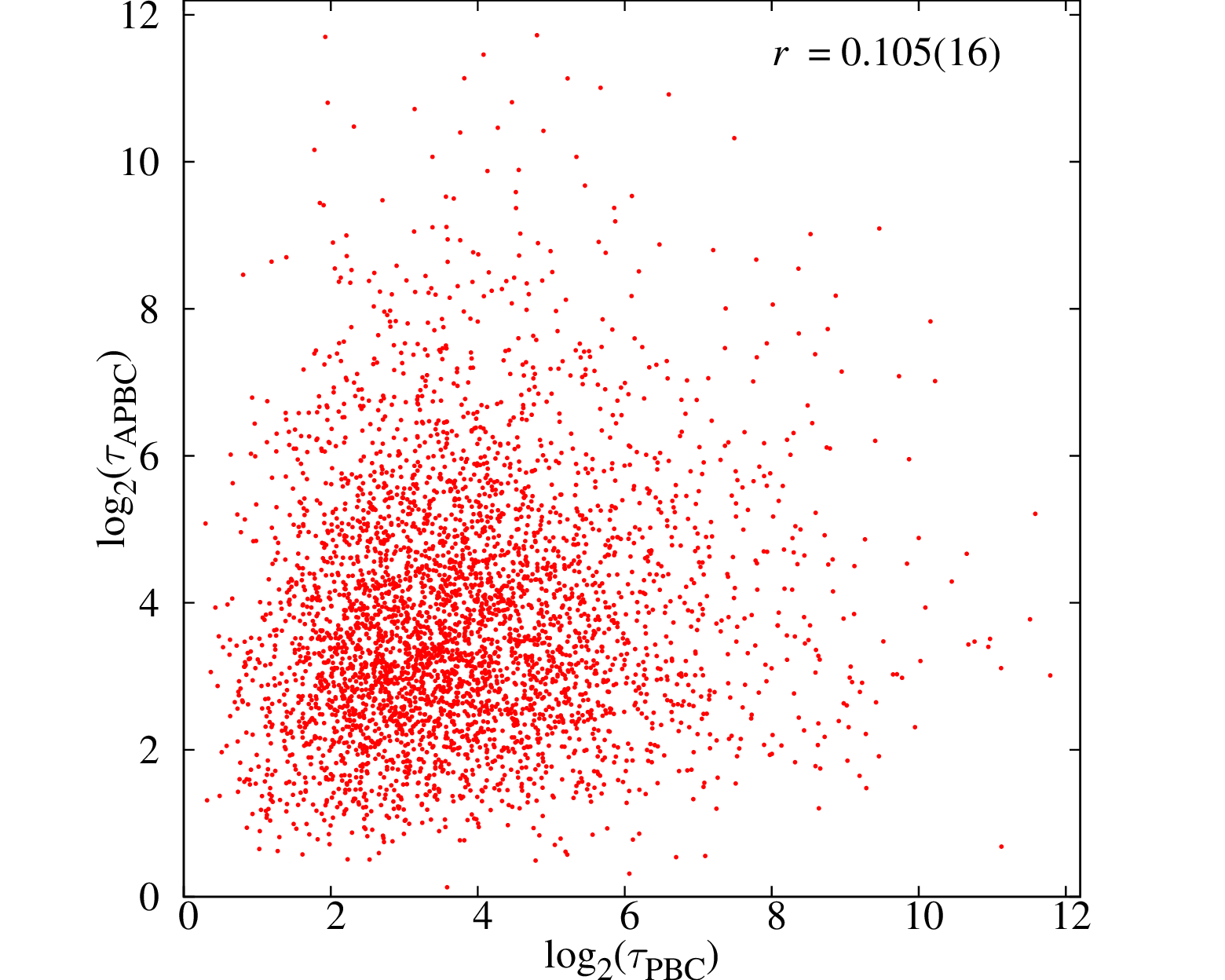}
\end{minipage}
\caption{\emph{The chaos integral and the temperature-mixing autocorrelation
time carry similar information.\/} We show here two scatter plots, obtained
from our $L=12$, $T_{\mathrm{min}}=0.414$ data.  {\bf Left:} We plot
$(\mathrm{log}_2(\mathrm{PBC}),I_{PBC})$ for each of our 4000 original PBC
samples.  {\bf Right:} for the temperature-mixing time $\tau$, we plot
$(\mathrm{log}_2(\mathrm{PBC}),\mathrm{log}_2(\mathrm{APBC}))$ for each of our
4000 sample pairs. All the $\tau$ are integrated autocorrelation times
(see~\cite{janus:10}) and are expressed in units of $N_\mathrm{mes} = 5000$
heat-bath steps (see Table~\ref{tab:parameters}).
 \label{fig:I-or-tau}}
\end{figure}
As we explained in Sect.~\ref{sect:T-chaos} the most appealing
numerical characterization of temperature is the auto-correlation time
$\tau$ for temperature-mixing along a parallel tempering
simulation~\cite{fernandez:13,martin-mayor:15}. Unfortunately, a
high-accuracy computation of $\tau$ is not a light task, so we need an
easier-to-compute alternative. A nice alternative is provided by the
chaos integral $I$ defined in Eq.~(\ref{eq:I})~\cite{fernandez:13}.

Indeed, see Fig.~\ref{fig:I-or-tau}---left, our estimations of $\tau$ and
$I$ are very strongly correlated. Furthermore, our main theme (namely
the very small correlation between the original PBC sample and its
APBC-transform) is maintained when we work in terms of $\tau$, see
Fig.~\ref{fig:I-or-tau}---right.

\section{Additional results}\label{ap:additional}

\begin{figure}[h]
\begin{minipage}[t]{\linewidth}
\includegraphics[height=0.525\linewidth,angle=270]{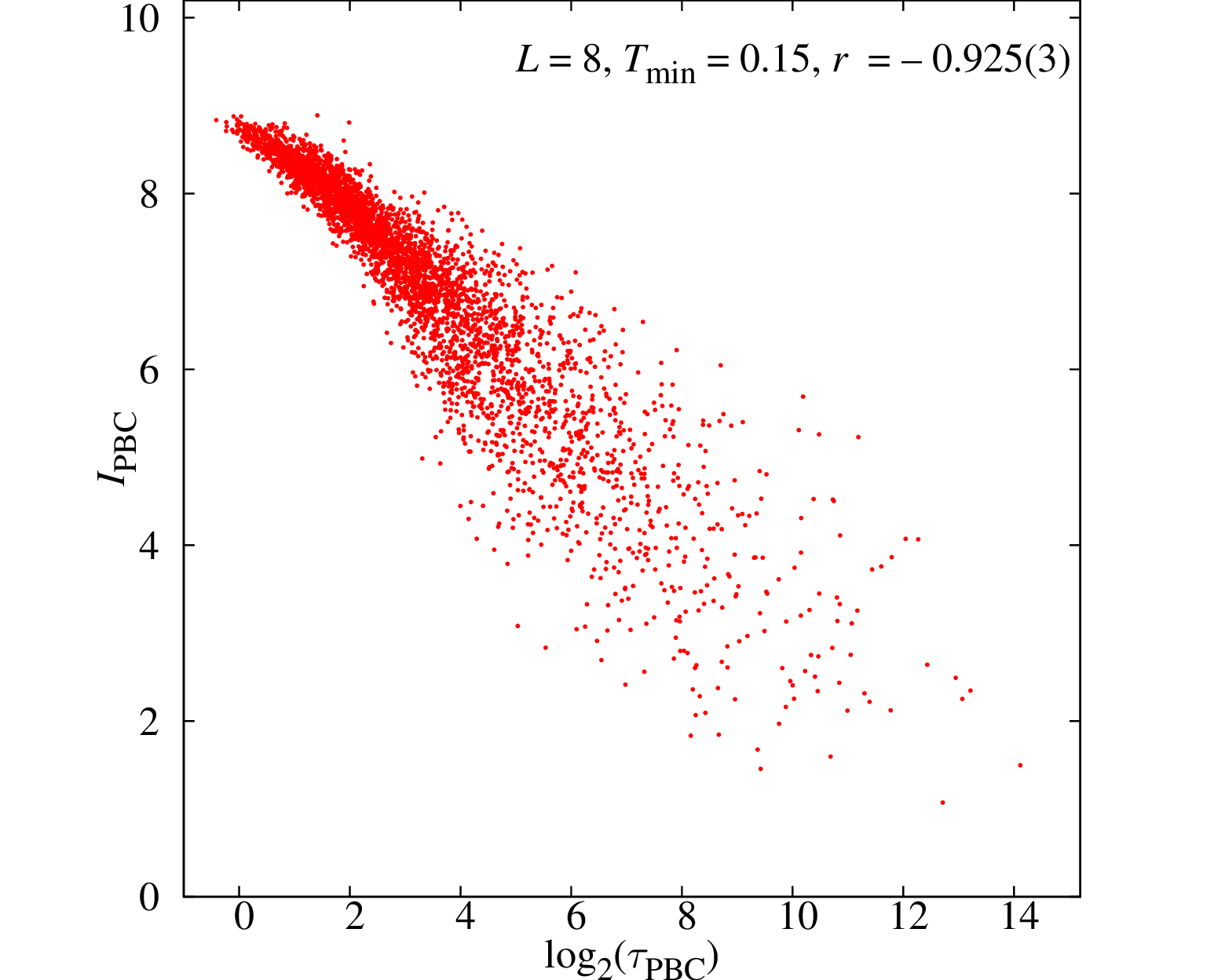}
\includegraphics[height=0.525\linewidth,angle=270]{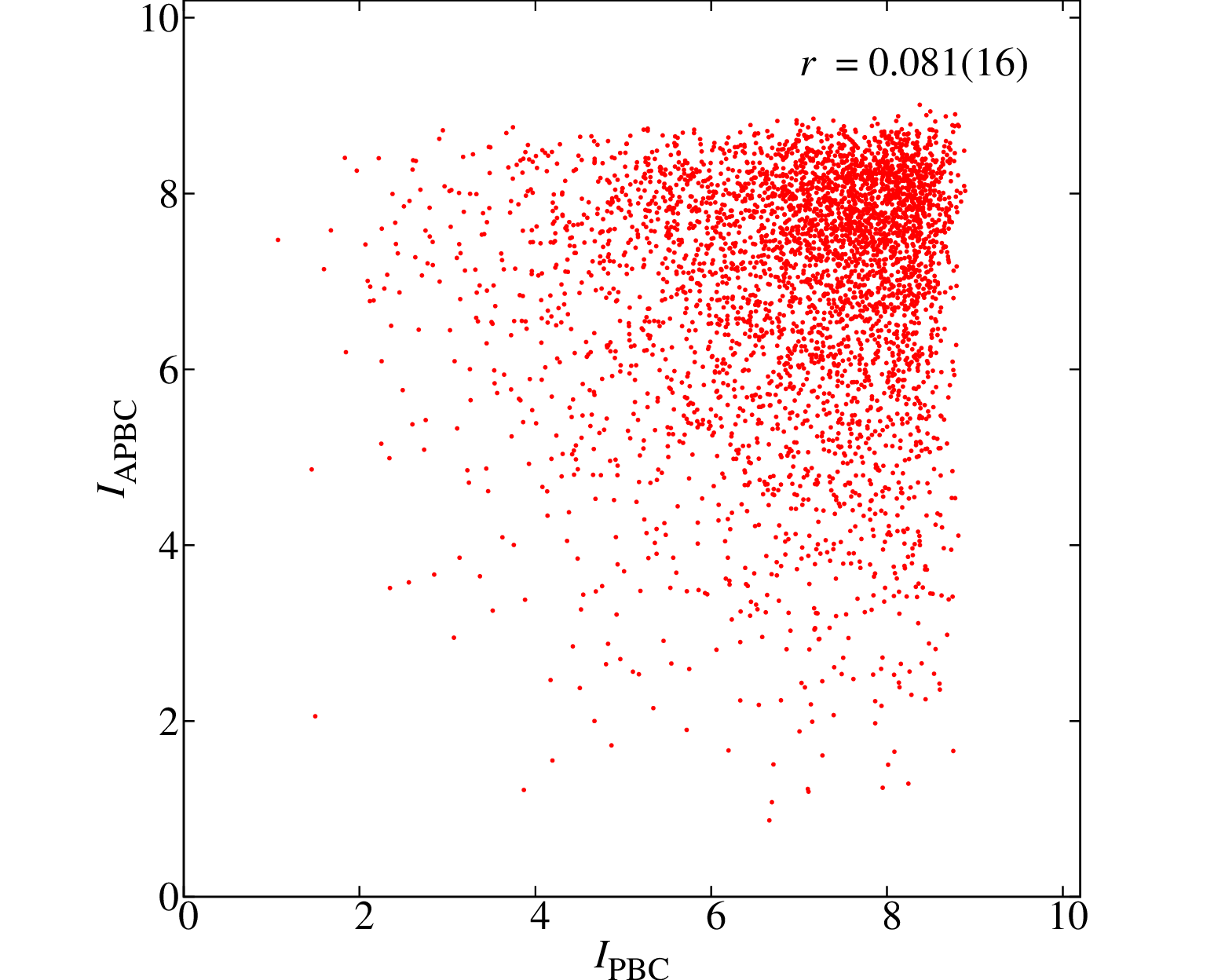}
\end{minipage}
\caption{We show two scatter plots, as obtained from our $L=8$ and
  $T_{\mathrm{min}}=0.15$ data.  {\bf Left:} The analogous of
  Fig.~\ref{fig:I-or-tau}---left. {\bf Right:} The analogous of
  Fig.~\ref{fig:scatter-I}.
 \label{fig:I-tau-I-I-L8}}
\end{figure}

\begin{figure}[h]
\begin{center}
\includegraphics[height=0.7\linewidth,angle=270]{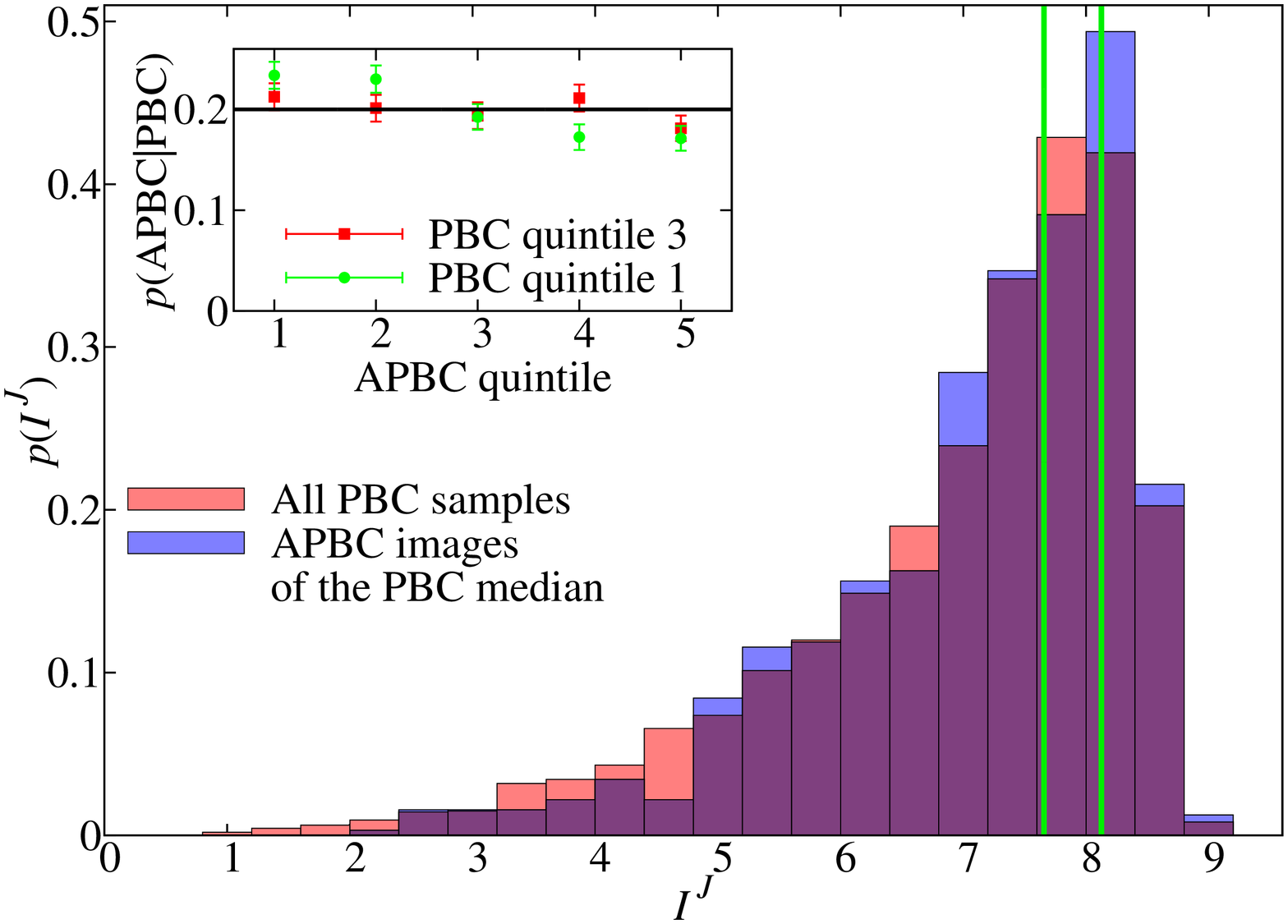}
\end{center}
\caption{The analogous of Fig~\ref{fig:quintiles}, as obtained from our $L=8$ and $T_{\mathrm{min}}=0.15$ data.
 \label{fig:quintiles-L8}}
\end{figure}

\begin{figure}[h]
\begin{minipage}[t]{\linewidth}
\includegraphics[height=0.525\linewidth,angle=270]{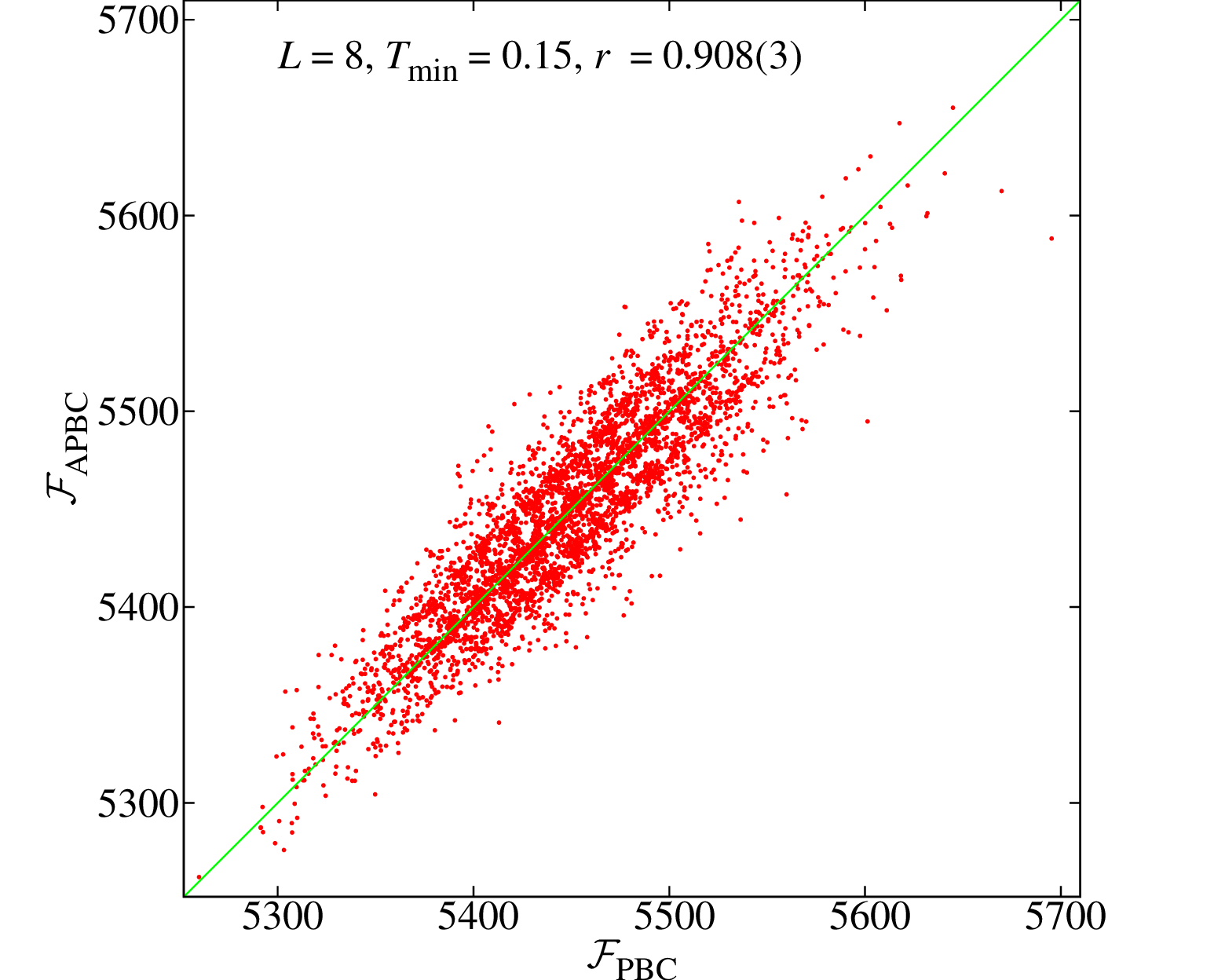}
\includegraphics[height=0.525\linewidth,angle=270]{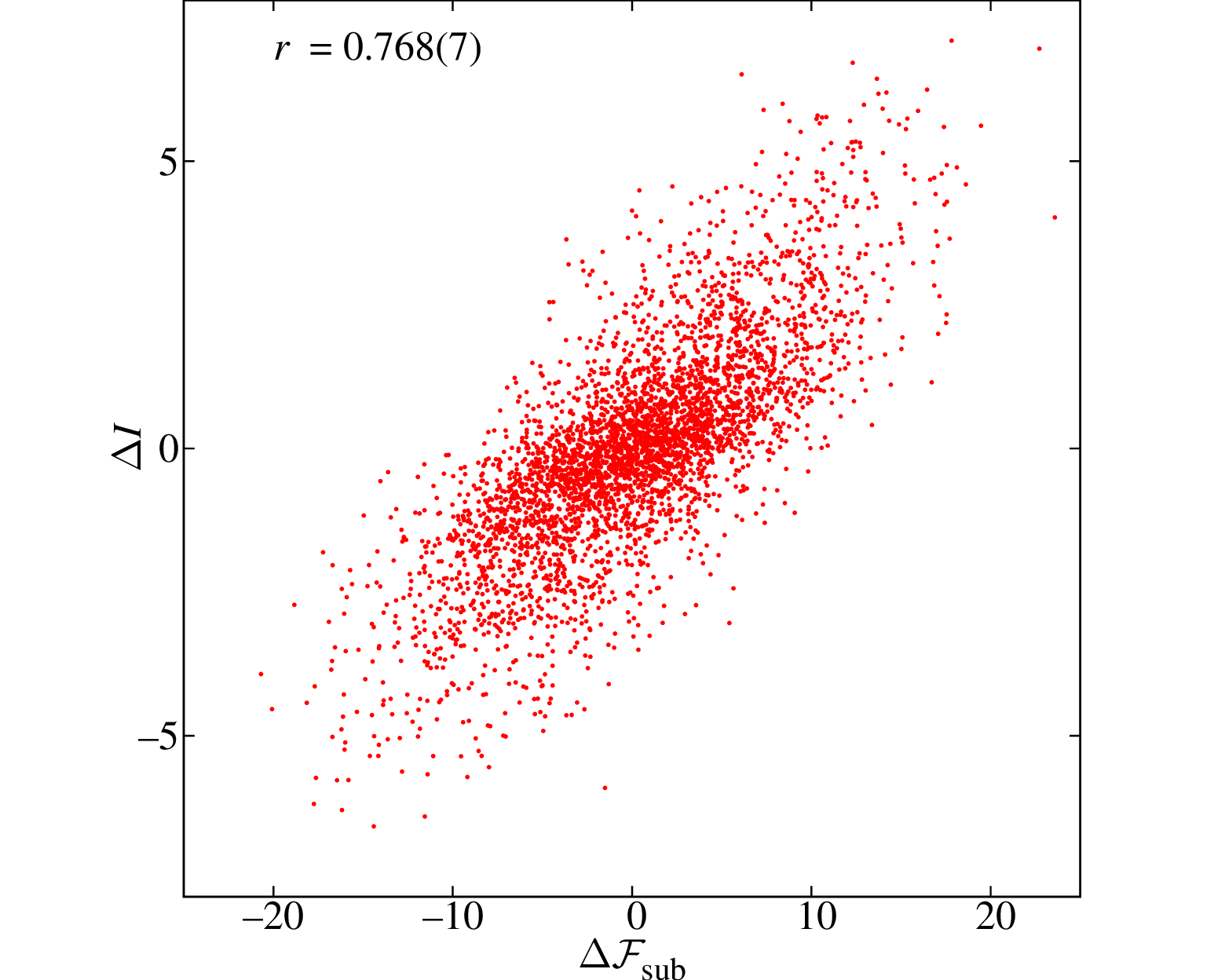}
\end{minipage}
\caption{We show two
scatter plots for our $L=8$ and $T_{\mathrm{min}}=0.15$ data (this figure
is the analogous of Fig~\ref{fig:F-I}, as obtained for these $L$ and $T_{\mathrm{min}}$). {\bf Left:} For the free-energy~(\ref{eq:F}) we plot $({\cal F}_\mathrm{PBC},{\cal F}_\mathrm{APBC})$ for each of our
4000 sample pairs. The green straight line is $y=x$. {\bf Right:} we plot
$({\cal F}_\mathrm{sub,PBC}-{\cal F}_\mathrm{sub,APBC},I_\mathrm{PBC}-I_\mathrm{APBC})$. 
 \label{fig:F-L8}}
\end{figure}

The purpose of this section is to show that neither the choice of
temperature interval nor of studied system size is critical. This
is evinced in Figs.~\ref{fig:I-tau-I-I-L8},~\ref{fig:quintiles-L8}
and~\ref{fig:F-L8}.

\section{A geometric inequality on correlations}\label{ap:ineq}
The assertion that the pair of stochastic variables
$(I_\mathrm{APBC},I_\mathrm{PBC})$ are essentially uncorrelated might
be surprising on the view of the mild correlations for $({\cal
  F}_\mathrm{PBC},I_\mathrm{PBC})$ [or, equivalently, $({\cal
    F}_\mathrm{APBC},I_\mathrm{APBC})$] and the very strong
correlations depicted in Fig. 4 for $({\cal F}_\mathrm{APBC},{\cal
  F}_\mathrm{PBC})$. A simple geometric argument explains how
misleading this way of reasoning might be. We thank one of our
referees for calling our attention to this issue.

We shall first obtain an inequality, and then apply it to our
problem. We start by considering a triplet of stochastic variables
$(X_0,X_1,X_2)$. Let $E(\cdots)$ denote the expectation value. For
each $X_i$ we define a related quantity $x_i$:
\begin{equation}
X_i=E(X_i)+\sigma_{ii}^{1/2} x_i\,,
\end{equation}
where $\sigma_{ii}$ is the variance of $X_i$. We note that the $x_i$
are normalized, in the sense that $E(x_i^2)=1$, that $E(x_i)=0$ and
that the correlation coefficient can be written as
\begin{equation}
r_{ij}=E(x_i\,x_j)\,.
\end{equation}
Now, we split the stochastic variable $x_i$ for $i=1,2$ as
\begin{equation}
x_i=r_{0i}\, x_0+ \tilde x_i\,.
\end{equation}
Note that
\begin{equation}
E(x_0\, \tilde x_i)= 0\ ,\quad E(\tilde x_i^2)=1-r_{0i}^2\,.
\end{equation}
It follows that
\begin{equation}
r_{12}=E(x_1\,x_2)=r_{01}r_{02}+E(\tilde x_1\, \tilde x_2)\,
\end{equation}
Finally, we recall that the Cauchy-Schwarz-Bunyakovsky inequality unfortunately
only implies $|E(x_1\, x_2)|\leq \sqrt{E(x_1^2) E(x_2^2)}$. Hence, the
most we can tell about  $r_{12}$ judging from $r_{01}$ and $r_{02}$ is
\begin{equation}\label{eq-appendix:ineq-geometrica}
r_{01}r_{02}-\sqrt{1-r_{01}^2}\sqrt{1-r_{02}^2} \ \leq\ r_{12}\ \leq\ r_{01}r_{02}+\sqrt{1-r_{01}^2}\sqrt{1-r_{02}^2}\,.
\end{equation}

In our case, the variables of interest are $X_1=I_\mathrm{PBC}$ and
$X_2=I_\mathrm{APBC}$. As for $X_0$ we can choose either ${\cal
  F}_\mathrm{PBC}$ or ${\cal F}_\mathrm{APBC}$ (these two quantities
  are so correlated that we can consider the most favourable case in
  which we identify them). Note that, in this approximation,
  $r_{01}=r_{02}\equiv r$. Hence, only for $r > 1/\sqrt{2}\approx 0.71$
  (much larger than the correlation we found), the
  inequality~(\ref{eq-appendix:ineq-geometrica}) guarantees some
  correlation, i.e., $r_{12}>0$.

\clearpage

\section*{References}
\providecommand{\newblock}{}

\end{document}